\documentclass[twocolumn,preprintnumbers,prl,aps,superscriptaddress]{revtex4}

\usepackage{latexsym}          
\usepackage[dvips]{graphicx}   
\usepackage{amsmath}           

\usepackage{amsfonts}          
\usepackage{amssymb}           

\usepackage[english]{babel}

\newtheorem{theorem}{Theorem}

\newcommand{\be}{\begin{equation}}
\newcommand{\ee}{\end{equation}}

\newcommand{\ef}[1]{\, #1}

\def\psibar{{\bar{\psi}}}

\def\smfrac#1#2{{\textstyle\frac{#1}{#2}}}
\newcommand{\smbinom}[2]{\genfrac{(}{)}{0pt}{1}{#1}{#2}}

\newcommand{\idf}{\mathit{Id}_{\rm f}}
\newcommand{\idr}{\mathit{Id}_{\rm r}}
\newcommand{\idrodd}{\hat{\mathit{Id}}_{\rm r}}
\newcommand{\bout}{b^{-}}
\newcommand{\bin}{b^{+}}

\newcommand{\tildina}{\raisebox{-3pt}{$\tilde{\phantom{\cdot}}$}}
\newcommand{\lui}{\ \raisebox{2pt}{\rule{15pt}{0.5pt}},\ }

\begin{document}

\title{Explicit characterization of the identity configuration \\ in an
Abelian Sandpile Model}

\author{Sergio Caracciolo}
\affiliation{Dip.~Fisica,
  Universit\`a degli Studi di Milano and INFN, via G.~Celoria 16, 20133 Milano, Italy}
\author{Guglielmo Paoletti}
\affiliation{Dip.~Fisica,
  Universit\`a degli Studi di Pisa and INFN, largo B.~Pontecorvo 3, 56127 Pisa, Italy}
\author{Andrea Sportiello}
\affiliation{Dip.~Fisica,
  Universit\`a degli Studi di Milano and INFN, via G.~Celoria 16, 20133 Milano, Italy}

\date{\today}

\begin{abstract}
\noindent
Since the work of Creutz, identifying the group identities for the
Abelian Sandpile Model (ASM) on a given lattice is a puzzling issue:
on rectangular portions of $\mathbb{Z}^2$ complex quasi--self-similar
structures arise.  We study the ASM on the square lattice, in
different geometries, and a variant with directed edges.  Cylinders,
through their extra symmetry, allow an easy determination of the
identity, which is a homogeneous function.  The directed variant on
square geometry shows a remarkable exact structure, asymptotically
self-similar.
\end{abstract}

\maketitle

\section{Introduction}

\noindent
Bak, Tang and Wiesenfeld (BTW) introduced the Abelian Sand\-pile Model (with a
different name) in \cite{btw} as a simple model for self-organized
criticality. The model has been soon generalized from the square lattice to
arbitrary graphs~\cite{dhar}.
It is a lattice automata, described in
terms of local height variables $z_i \in \mathbb{N}$, i.e.~the
number of ``particles'' at site $i$, so that a configuration
is given by set $C = \{ z_i \}$.
Particles fall into the system
through a random memoryless process, and, as soon as some height $z_i$
is higher than the local threshold value $\bar{z}_i$, it
\emph{topples} to the neighbouring sites, i.e.~the value $z_i$
decreases by the (integer) amount 
$\Delta_{ii}$, while, for each site $j \neq i$, increases by the
(integer) amount $-\Delta_{ij}$. This relaxation process (the
\emph{avalanche}) is guaranteed to stop in a finite number of steps if
$\bout_i := \sum_j \Delta_{ij} \geq 0$ for all sites $i$, and
a certain technical condition
is satisfied (the matrix $\Delta$ must be a 
\emph{strictly dissipative} diffusion kernel).
All the heights remain non-negative if this is the case in
the starting configuration, and $\bar{z}_i \geq \Delta_{ii}$.
For future convenience we also define
$\bin_i := \sum_j \Delta_{ij}^{T}$, and assume that also these
quantities be non-negative.

A graph, in general directed and with multiple edges, is thus defined
through the non-diagonal part of $-\Delta$, seen as an adjacency
matrix, while the non-zero values 
$b^{\pm}_i$ are regarded as (in- or out-coming) ``connections
to the border'' (and the sites $i$ with $\bin_i$ or $\bout_i$ non-zero
are said to be ``on
the border'').  If $\Delta$ is symmetric, we have an undirected graph,
and automatically $\bin_i = \bout_i$
We just write $b_i$ for $\bin_i$ and $\bout_i$ when clear that they
are equal.

This is the case in the original BTW model, which is defined on a
portion of the $\mathbb{Z}^2$ lattice. We have in this case $\bar{z}_i =
\Delta_{ii} = 4$ on each site, and $\Delta_{ij}=-1$ on all pairs of
neighbouring sites. If we are on a rectangular domain, say of sides
$L_x$ and $L_y$, we have $b_i=1$ for vertices on the sides of the
rectangle, and $b_i=2$ for the corners.

In \cite{btw} it is described how, already for this case, the avalanches show
an unpredictable dynamics, with power-law size distribution,
which candidated the model as a toy description of many interesting
features, such as the $1/f$ noise.

Afterwords, Deepak Dhar and collaborators, in a series of relevant works
\cite{dhar, majdhar, dhar_alg}, elucidated a deep structure
of the Markov chain related to the general ASM:
the set of \emph{stable} configurations 
$S=\bigotimes_i \{0, 1, \ldots, \bar{z}_i-1 \}$, the ones where no
toppling can occur, is
divided into a set $T$ of \emph{transient} and a set $R$ of
\emph{recurrent} ones, the latter being visited an
infinite number of times by the chain, and forming a single ergodic
basin. Given the natural partial ordering, 
$C \preccurlyeq C'$ iff $z_i \leq z'_i$ for all $i$,
then $R$ is in a sense ``higher'' than $T$, more precisely
\begin{align}
\nexists \  (C,C') \in T \times R \,:                && C \succ C'
\ef;
\\
\forall  \, C \in T \quad \exists\, C' \in R \,: && C \prec C'
\ef.
\end{align}
In particular, the \emph{maximally-filled} configuration 
$C_{\rm max} = \{ \bar{z}_i -1 \}$ is in $R$, and higher than any
other stable configuration.

The set $R$ has an underlying abelian structure, for
which a presentation is explicitly constructed in terms of the
matrix $\Delta$, through the (heavy) study of its
Smith normal form. Shortcuts of the construction and more explicit
analytical results are achieved in the special case of a rectangular
$L_x \times L_y$ portion of the square lattice, and still stronger
results are obtained for the case of $L_x = L_y$~\cite{dhar_alg}.

Call $\tilde{a}_i$ the operator which adds a particle at site $i$ to a
configuration $C$, and $a_i$ the formal operator which
applies $\tilde{a}_i$,
followed by a sequence of topplings
which makes the configuration stable. Remarkably, the final
configuration $a_i C$ is independent from the sequence of topplings,
and also, applying two operators, the two configurations 
$a_j a_i C$ and $a_i a_j C$ coincide, so that at a formal level
$a_i$ and $a_j$ do commute.

More precisely, if $a_i$
acting on $C$ consists of the fall of a particle in $i$,
$\tilde{a}_i$, and the sequence of topplings 
$t_{i_1}, \ldots, t_{i_k}$ on sites $i_1, \ldots, i_k$, 
the univocal definition of $a_i$ and the
commutation of $a_i$ and $a_j$ follow from the two facts:
\begin{align}
z_j \geq \bar{z}_j: &&
t_j \tilde{a}_i C  
&= \tilde{a}_i t_j C
\ef;
&&
\\
z_{i} \geq \bar{z}_{i} \textrm{\ and\ }
z_{j} \geq \bar{z}_{j}:&&
t_i t_j C 
&= t_j t_i C
\ef.
\end{align}
Another consequence is that,
instead of doing all the topplings immediately, we can
postpone some of them after the following $\tilde{a}$'s, and still get
the same result. Similar
manipulations show that the relation
\be
\label{eq.a4}
a_i^{\Delta_{ii}} = \prod_{j \neq i} a_j^{- \Delta_{ij}}
\ee
holds when applied to an arbitrary configuration.

These facts lead to the definition of an abelian
semi-group operation between
two configurations, as the sum of the local height variables $z_i$,
followed by a relaxation process~\cite{creutz}:
\be
\label{eq.gprod}
C \oplus C' = 
\bigg( \prod_i a_i^{z_i} \bigg) C' =
\bigg( \prod_i a_i^{z'_i} \bigg) C
\ef.
\ee
For a configuration $C$, we define multiplication by a positive
integer:
\be
k\, C
=
\underbrace{C \oplus \cdots \oplus C}_k
\ef.
\ee
The set $R$ of recurrent configurations is special, as each operator
$a_i$ has an inverse in this set, so the operation above, restricted
to $R$, is raised to a group operation.  According to the Fundamental
Theorem of Finite Abelian Groups, any such group must be a ``discrete
thorus'' $\mathbb{Z}_{d_1} \times \mathbb{Z}_{d_2} \times \cdots
\times \mathbb{Z}_{d_g}$, for some integers $d_1 \geq d_2 \geq \cdots
\geq d_g$, and such that $d_{\alpha +1}$ divides $d_{\alpha}$ for each
$\alpha=1, \ldots, g-1$.  The values $d_{\alpha}$, called
\emph{elementary divisors} of $\Delta$, and a set of generators
$e_{\alpha}$ with the proper periodicities, can be constructed through
the normal form decomposition~\cite{dhar_alg}.  The composition of
whatever $C=\{ z_i \}$
with the set $R$ acts
then as a translation on this thoroidal geometry. A further
consequence is that, for any recurrent configuration $C$, the inverse
configuration $(-C)$ is defined, so that $k\,C$ is defined for $k \in
\mathbb{Z}$.

Consider the product over sites $i$ of equations~(\ref{eq.a4})
\be
\label{eq.653675}
\prod_i a_i^{\Delta_{ii}} = 
\prod_i \prod_{j \neq i} a_j^{- \Delta_{ij}}
= \prod_i a_i^{-\sum_{j \neq i} \Delta_{ji}}
\ee
On the set $R$, the
inverses of the formal operators $a_i$ are defined, so that we can
simplify common factors in (\ref{eq.653675}), recognize the expression
for $\bin$, and get
\be
\label{eq.aid}
\prod_i a_i^{\bin_i} = I
\ee
so that $\prod_i a_i^{\bin_i} C = C$ is a necessary condition for $C$ to
be recurrent (but it is also sufficient, as no transient configuration
is found twice in the same realization of the Markov chain),
and goes under the name of \emph{identity test}. 

This condition turns into an equivalent and
com\-pu\-ta\-tionally-\-cheaper procedure, called \emph{burning test},
of which a side product, in case of positive answer, is a spanning
arborescence rooted on the vertices of the border. So, the burning
test provides us a bijection between the two ensembles, of recurrent
sandpile configurations and rooted arborescences with roots on the
border. This is in agreement with Kirchhoff Matrix-Tree theorem, which
states that the number of such arborescences is given by $\det
\Delta$, while the number of recurrent configurations is known to be
$\det \Delta$ as the first step of the procedure which determines the
elementary divisors~$d_{\alpha}$~\cite{majdhar}.

If, for the graph identified by $\Delta$, a planar embedding exists,
with all sites $i$ such that $\bin_i > 0$ on the most external face,
then the planar dual of a rooted arborescence coming from the burning
test is a spanning tree on the planar-dual graph.

\begin{figure}
 \centering
\includegraphics[scale=1.1]{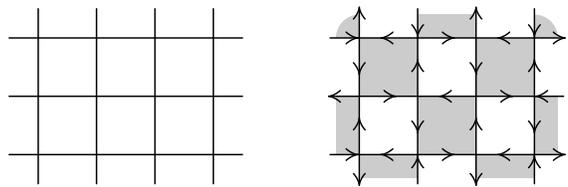}
 \caption{\label{fig.1234}%
Left: a portion of the square lattice. Right: a portion of the
directed square lattice we considered in this work, with in- and
out-edges alternated cyclically, and white and gray faces with arrows
oriented clockwise and counter-clockwise.}
\end{figure}

This is clear for the undirected case $\Delta = \Delta^{T}$, and needs
no more words.  If the graph is directed, the arborescence is directed
in the natural way, while the dual tree is constrained in some
complicated way (some combinations of edge-occupations are forced to
fixed values). However, the graphical picture simplifies if, on the
planar embedding, in-coming and out-coming edges are cyclically
alternating (and all plaquettes have consistent clockwise or
counter-clockwise perimeters), and all sites with either $\bin$ or $\bout$
positive are on the most external face, with in- and out-going arrows
cyclically alternating (cfr.~for example Figure~\ref{fig.1234}).
We will call a directed graph of this kind a
\emph{planar alternating directed graph}.

The connection with uniform spanning trees and the Kirchhoff theorem
explains \emph{a posteriori} the arising of self-organized
criticality, i.e.~the appearence of long-range behaviour with no need
of tuning any parameter. Indeed, uniform spanning trees on regular
2-dimensional lattices are a $c=-2$ logarithmic 
Conformal Field Theory (CFT), 
and have no parameter at all, being a peculiar limit $q \to 0$ of the
Potts model in Fortuin-Kasteleyn formulation~\cite{alan_bcc}, or a
limit of zero curvature in the $\mathrm{OSP}(1|2)$ non-linear
$\sigma$-model \cite{noi}. If instead one considers the larger
ensemble of spanning forests, in a parameter $t$ counting the
components (or describing the curvature of the $\mathrm{OSP}(1|2)$
supersphere), the theory in two dimensions is scale-invariant for
three values: at $t=0$ (the spanning trees, or the endpoint of the
ferromagnetic critical line of Potts), at the infinite-temperature
point $t= \infty$, and at some non-universal negative $t$
corresponding to the endpoint of the anti-ferromagnetic critical line
of Potts, being $t_c = -1/4$ on the square lattice.  Through the
correspondence with the non-linear $\sigma$-model, one can deduce at a
perturbative level the renormalization group flow, and, in particular,
that the system is asymptotically free for $t>0$~\cite{noi, conclaudia}, so that $t=0^+$ is an ultraviolet fixed point.

An interesting connection between the height variables in the ASM and
the fields of the underlying CFT is performed 
in \cite{majdhar, priez, ruold, ruepir_new}, and especially
in~\cite{ruepir_new}.

\section{Identities}

\begin{figure}
 \centering
\includegraphics[scale=0.6]{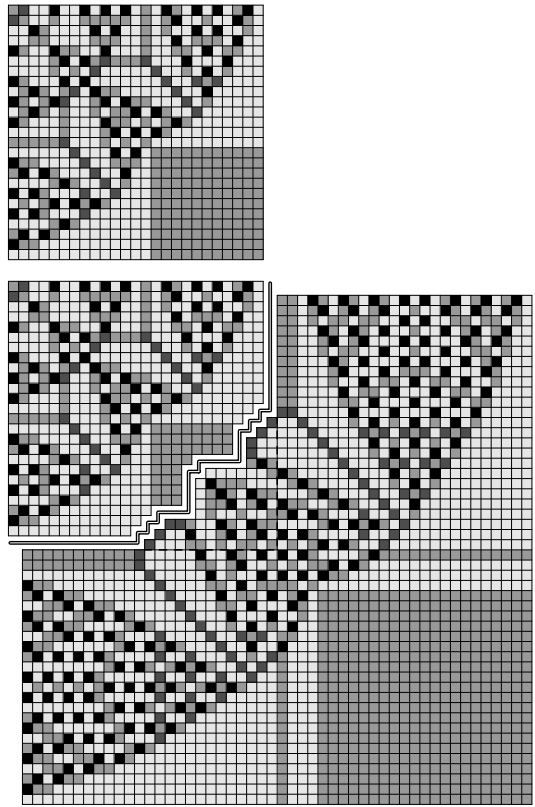}
 \caption{\label{fig.idsASM}%
The top-left corners of the identities $\idr^{(L)}$ for $L=50$ and
$L=100$ in the BTW model (the other quadrants are related by
symmetry). Heights from 0 to 3 correspond to gray tones from dark to
light.  The smaller-size identity is partially reproduced at the
corner of the larger one, in a fashion which resembles the results of
Theorem~\ref{theorem1}.}
\end{figure}

\noindent
Given the algebraic relation (\ref{eq.aid}), and the semi-group
operation (\ref{eq.gprod}), one could define the \emph{frame}
configuration $\idf$ as the one with $z_i = \bin_i$ for all $i$, and
realize that it acts as an identity on recurrent
configurations, $\idf \oplus C = C$ if $C \in R$.
Conversely, in general it does not leave unchanged a
transient configuration, and in particular, as, for any relevant
extensive graph, $\idf$ is itself transient, we have that $\idf \oplus
\idf$, and $\idf \oplus \idf \oplus \idf$, and so on, are all
different, up to some number of repetitions $k$ at which the
configuration is sufficiently filled up with particles to be
recurrent. We call $\idr$ this configuration, and $k(\Delta)$ the
minimum number of repetitions of $\idf$ required
in the ASM identified by $\Delta$ (we name it the \emph{filling
number} of $\Delta$).  The configuration $\idr$
is the identity in the abelian group $\mathbb{Z}_{d_1} \times \cdots
\times \mathbb{Z}_{d_g}$ described above, and together with a set of
generators $e_{\alpha}$, completely identifies in a constructive way
the group structure of the statistical ensemble. The relevance of this
configuration has been stressed first by M.~Creutz~\cite{creutz}, so
that we shall call it \emph{Creutz identity}.
See fig.~\ref{fig.idsASM} for an example.

Unfortunately, despite many efforts, it has not been possible to give
a closed-formula recipe for this identity state on given large
lattices, not even in the case of a $L \times L$ square, and the
direct numerical investigation of these configurations at various
sizes has produced peculiar puzzling pictures \cite{creutztoys}.

In a large-side limit, we have the formation of curvilinear triangular
regions of extensive size (of order $L$), filled with regular
patterns, and
occasionally crossed by straigth ``defect lines'', of widths of order
1, which, furthermore, occasionally meet at Y-shaped ``scattering
points'', satisfying peculiar conservation laws~\cite{us_pseudoprop}.

\begin{table}
\raisebox{-45pt}{\rule{0pt}{100pt}}%
\begin{tabular}{r|ccccccccccc|l}
\hline
\hline
$L$    & 2 & 4 & 6 & 8 & 10 & 12 & 14 & 16 & 18 & 20 & 22 &
$\scriptstyle{\vartriangleright}$ \\
\cline{1-12}
$k_L$ & 1 & 4 & 7 & 13 & 19 & 27 & 35 & 46 & 58 & 71 & 87 &
$\scriptstyle{\blacktriangleright}$ \\
\cline{1-12}
\rule{0pt}{12pt}%
$\scriptstyle{\vartriangleright}$
& 24 & 26 & 28 & 30 & 32 & 34 & 36 & 38 & 40 & 42 & 44 &
$\scriptstyle{\vartriangleright}$ \\
\cline{2-12}
$\scriptstyle{\blacktriangleright}$
& 103 & 119 & 138 & 156 & 180 & 198 & 226 & 248 & 276 & 305 & 334 &
$\scriptstyle{\blacktriangleright}$ \\
\cline{2-12}
\rule{0pt}{12pt}%
$\scriptstyle{\vartriangleright}$
& 46 & 48 & 50 & 52 & 54 & 56 & 58 & 60 & 62 & 64 & \\
\cline{2-12}
$\scriptstyle{\blacktriangleright}$
& 367 & 397 & 430 & 464 & 499 & 538 & 572 & 615 & 653 & 699 & \\
\hline
\hline
\end{tabular}
\caption{\label{tab.kL}Values of $k_L$ for the BTK Abelian Sandpile on
  square geometries of even size, for $L=2, \ldots, 64$.}
\end{table}

Similar features emerge also for the filling numbers, e.g.~on
the square lattice of size $L$, the index $k_L \equiv k(\Delta^{(L)})$
is not badly fitted, for even $L$, by a parabola 
$k_L \simeq L^2/6 + o(L^2)$, but showing fluctuations due to unpredictable
number-theoretical properties of~$L$. The challenging sequence of
these numbers, for $L$ up to 64, is given in Table~\ref{tab.kL}.
It should be noted that, conversely, odd sizes $2L+1$ are related to
$2L$ through a property proven in \cite[sec.~7]{dhar_alg}.


The determination of $\idr$ for a given graph is a
procedure polynomial in the size of the graph. E.g.~one could prove
that $k(\Delta)$ is sub-exponential, and that relaxing $C \oplus C'$
for $C$ and $C'$ both stable is polynomial, then one can produce
the powers $2^{s} \, \idf$ recursively in $s$, up to get twice the same
configuration. Better algorithms exist however, see for
example~\cite{creutz}.

Still, one would like to have a better understanding of these identity
configurations.
This motivates us to the study of the ASM on different regular graphs,
such that they resemble as much as possible the original model, but
simplify the problem in some regards so that the family (over $L$) of
resulting identities can be understood theoretically.

In this research, we keep in mind a few principles: (1)~we wish to
simplify the problem in some structural way; (2)~we want to preserve
the property of the $L \times L$ lattice, discussed in
\cite[sec.~4]{dhar_alg}, of reconstruction of the elementary divisors
of $\Delta$ from a suitable matrix $n_{y y'}$, outcome of a machinery
for producing an ``economic'' presentation of the group, in terms of
$\mathcal{O}(L)$ generators $a_i$ of the whole set of $L^2$;
(3)~possibly, we want to preserve planarity, and the interpretation of
dual spanning subgraphs as spanning trees on the dual lattice, i.e.~we
want to use either planar undirected graphs, or planar alternating
directed graphs (according to our definition above).


\section{A prolog: cylindric geometries}

\noindent
A first possible simplification comes from working on a cylindrical
geometry. We call respectively \emph{periodic}, \emph{open} and
\emph{closed} the three natural conditions on the boundaries of a 
$L_x \times L_y$ rectangle. 
For example, in the BTW Model, for site $(i,1)$, in the
three cases of periodic, open and closed boundary conditions on the
bottom horizontal side we would have the toppling rules
\begin{align*}
\textrm{periodic:}&&
z_{i,1} & \to z_{i,1} -4;
&
z_{i\pm 1,1} & \to z_{i\pm 1,1} +1;
\\
&&
z_{i,2} & \to z_{i,2} +1;
&
z_{i,L_y} & \to z_{i,L_y} +1;
\\
\textrm{open:}&&
z_{i,1} & \to z_{i,1} -4;
&
z_{i\pm 1,1} & \to z_{i\pm 1,1} +1;
\\
&&
z_{i,2} & \to z_{i,2} +1;
&&
\\
\textrm{closed:}&&
z_{i,1} & \to z_{i,1} -3;
&
z_{i\pm 1,1} & \to z_{i\pm 1,1} +1;
\\
&&
z_{i,2} & \to z_{i,2} +1;
&&
\end{align*}
The external face, with $b^{\pm}_i \neq 0$, is in correspondence of
open boundaries, so, as we want a single external face, if we take
periodic boundary conditions in one direction (say, along $x$), the
only possible choice in this framework is to take closed and open
conditions on the two sides in the $y$ direction.

Our notation is that $\bar{z}_i$ takes the same value everywhere: 
on open boundaries, $\bout_i$ and $\bin_i$ are determined
accordingly, while on closed boundaries either we add some extra
``loop'' edges, or we take $\bar{z}_i - \Delta_{ii} > 0$.

The cylindric geometry has all the non-trivial features of the
original ASM for what concerns group structures, polynomial bound on
the relaxation time in the group action, connection with spanning trees,
and so on (even the finite-size corrections to the continuum-limit CFT
are not more severe on a $L_x \times L_y$ cylinder than on a $L_x
\times L_y$ open-boundary rectangle), \emph{but}, for what concerns
the determination of $\idr$ through relaxation of $k \idf$, the system
behaves as a quasi-unidimensional one, and $\idr$ is in general
trivially determined.

\begin{figure}
 \centering
\includegraphics[scale=1]{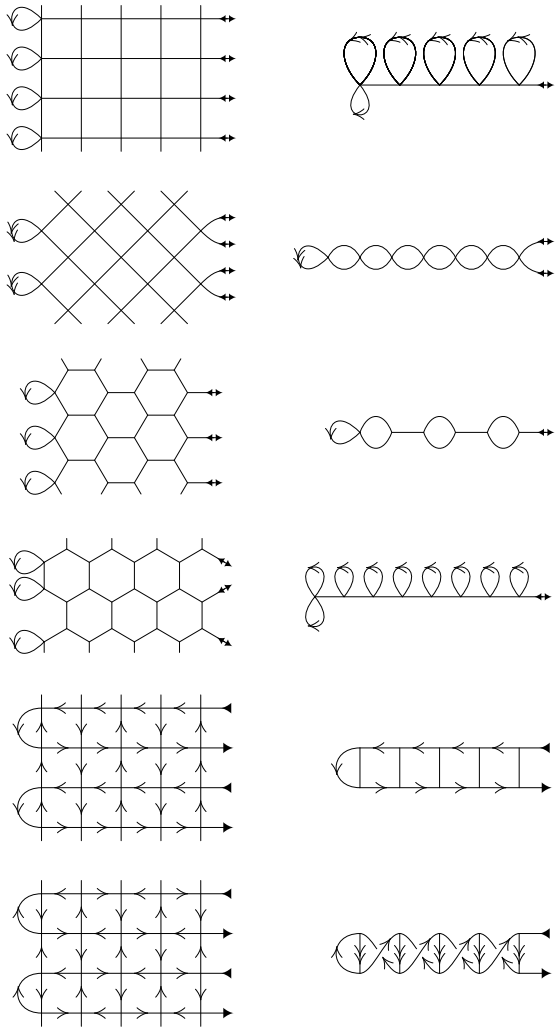} 
 \caption{\label{fig.quasi1D}%
Reduction to a quasi-1-dimensional system for the ASM on cylindric
geometry, on a few examples all having $\idr$ coinciding with 
$C_{\rm max}$.
From top to bottom: a portion of the square lattice, in the two
orientations; of the hexagonal lattice, in the two orientations, of
Manhattan and pseudo-Manhattan lattices.
Plain edges correspond to $\Delta_{ij} = \Delta_{ji} = -1$, while a
directed edge (from $i$ to $j$) with $k$ arrows correspond to
$\Delta_{ij}=-k$.
In-(out-)coming bold arrows correspond to $b^-$ ($b^+$)
equal to 1, while the
lozenge-shaped double-arrows
correspond to $b^-=b^+=1$. 
A loop with $k$ arrows on $i$ corresponds to
$\bar{z}_i-\Delta_{ii}=k$, otherwise $\bar{z}_i=\Delta_{ii}$.}
\end{figure}

Furthermore, in many cases
$\idr$ is just the maximally-filled configuration $C_{\rm max}$.
This fact is easily proven, either by checking
that $\idr \oplus \idf = \idr$, which is easy in the quasi-1-dimensional
formulation (or in other words, by exploiting the translation symmetry
in one direction), or with a simple burning-test argument, in
the cases in which on each site there is a single incoming
edge from sites nearer to the border.
Figure \ref{fig.quasi1D} shows a few examples.

Such a peculiar property has a desiderable consequence on the issue of
inversion of a recurrent configuration. Indeed, if $C$ and $C'$ are in
$R$, we have that
\be
(-C) \oplus C' = 
\idr \oplus (-C) \oplus C' =
(\idr -C) \oplus C'
\ee
where, if $C=\{z_i\}$, $(-C)=\{\tilde{z}_i\}$ is the inverse
configuration we seek, but $\idr -C$ is
simply the configuration with 
$\tilde{z}_i = \bar{z}_i - z_i - 1$, and
has all non-negative heights.

\section{Pseudo-Manhattan Lattice}


\noindent
From this section on, we will concentrate on square geometries, on the
square lattice with edges having a given orientation, and all vertices
having in- and out-degree equal to 2.


For this reason, we can work with $\bar{z}_i=2$, so that $S=\{0,1\}^V$
instead of $\{0,1,2,3\}^V$ (a kind of simplification, as from
``CMYK'' colour printing to ``black and white''). In this step we
lose in general a bit of symmetry: e.g.~on a square of size $2L$ we
have still the four rotations, but we lose the reflections, which are
arrow-reversing, while on a square of size $2L+1$ we lose rotations
of an angle $\pi/2$, and only have rotation of $\pi$.

Square lattices with oriented edges have already been considered in
Statistical Mechanics, especially in the variant called ``Manhattan
Lattice'' (i.e.~with horizontal edges oriented east- and west-bound
alternately on consecutive rows, and coherently within a row, and
similarly for vertical edges), cfr.~for example
\cite{capograssb}. However this lattice in two dimensions is not
a planar alternating directed graph, and the results for the related
ASM model will be discussed only briefly in last section.

We start by analysing a less common variant, which is better behaving
for what concerns the ASM model, and which we call
\emph{pseudo-Manhattan lattice} (it appears, for example, in the
totally unrelated paper \cite{Chalk}).  
In this case, the horizontal
edges are oriented east- and west-bound alternately in both directions
(i.e.~in a chequer design), and similarly for vertical edges. As a
result, all plaquettes have cyclically oriented edges.  A small
portion of this lattice is shown in Figure \ref{fig.1234}, where it is
depicted indeed as the prototype planar alternating directed graph.
Conventionally, in all our examples (unless otherwise specified) we
fix the orientations at the top-right corner to be as in the top-right
corner of Figure~\ref{fig.1234}.

Quite recently, in \cite{dharnew} both Manhattan and pseudo-Manhattan
lattices have been considered as an interesting variant of the ASM
model (the latter under the name of \emph{F-lattice}), with
motivations analogous to ours. This corroborates our claim that these
variants are natural simplifications of certain features in the
original BTK model.

\begin{figure}
 \centering
\includegraphics[scale=0.6]{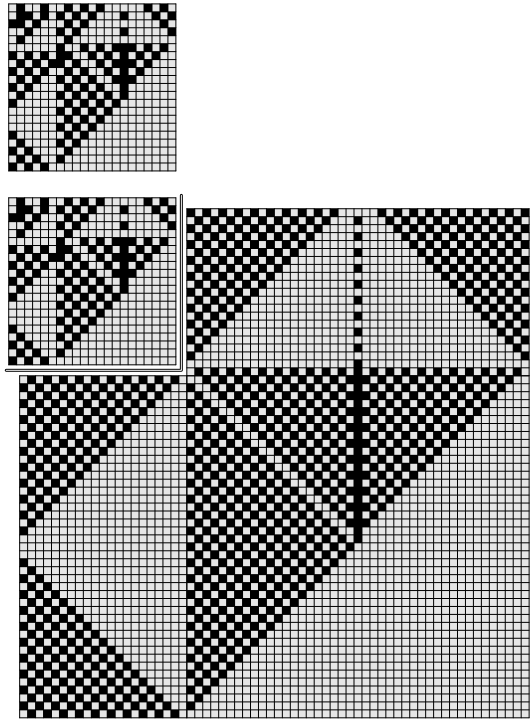}
 \caption{\label{fig.idsASMmanh}%
The top-right corners of the identities $\idr^{(L)}$ for $L=21$ and
$L=64$ (remark: $64 = 3 \cdot 21 + 1$).
The smaller-size one is exactly reproduced at the corner of the
larger one, while the rest of the latter has an evident regular
structure, according to theorem~\ref{theorem1}.}
\end{figure}

The main feature of the Creutz identities, on square portions of the
pseudo-Manhattan lattice with even side, is self-similarity for sides
best approximating the ratio $1/3$, up to a trivial part, as
illustrated in Figure~\ref{fig.idsASMmanh}.

The precise statement is in the following Theorem \ref{theorem1}.
Call $\idr^{(L)}$ the set of heights in the Creutz identity for the
square of side $2L$, encoded as a $L \times L$ matrix for one of the
four portions related by rotation symmetry. Say that index
$(1,1)$ is at a corner of the lattice, and index $(L,L)$ is in the
middle. We have

\begin{theorem}
\label{theorem1}
Say $L=3 \ell + s$, with $s=1,2,3$. Then, $\idr^{(L)}$ is determined
from $\idr^{(\ell)}$ and a closed formula, and thus, recursively,
by a deterministic telescopic procedure in at most 
$\lfloor \log_3 L \rfloor$ steps.

For $s=1$ or $3$,
we have $\big( \idr^{(L)} \big)_{ij} = \big( \idr^{(\ell)} \big)_{ij}$
if $i,j \leq \ell$,
otherwise 
$\big( \idr^{(L)} \big)_{ij} = 0$ iff, for $i+j+s$ even,
\begin{subequations}
\label{eq.conds}
\begin{align}
i &< \ell,       &    |L-\ell-j|-i  &> 0;            \\
j &> \ell,       &    |2\ell+1-i|+j &< 2 \ell + s;   \\
i &\leq 2 \ell,  &                j &= L-\ell;
\end{align}
and, for $i+j+s$ odd,
\begin{align}
i &> \ell,       &    |L-\ell-i|-j  &> 0;            \\
i &> \ell - 1,   &    |2 \ell + s -j|+i &< 2\ell;    
\end{align}
\end{subequations}

If $s=2$ the same holds with $i$ and $j$ transposed in
$\idr^{(L)}$ and $\idr^{(\ell)}$ (but not in (\ref{eq.conds})).
Furthermore, $k_L=\frac{L(L+1)}{2}$.
\end{theorem}

\begin{figure*}
\setlength{\unitlength}{55pt}
\begin{picture}(8,2.75)
\put(0,0){\includegraphics[scale=1.1]{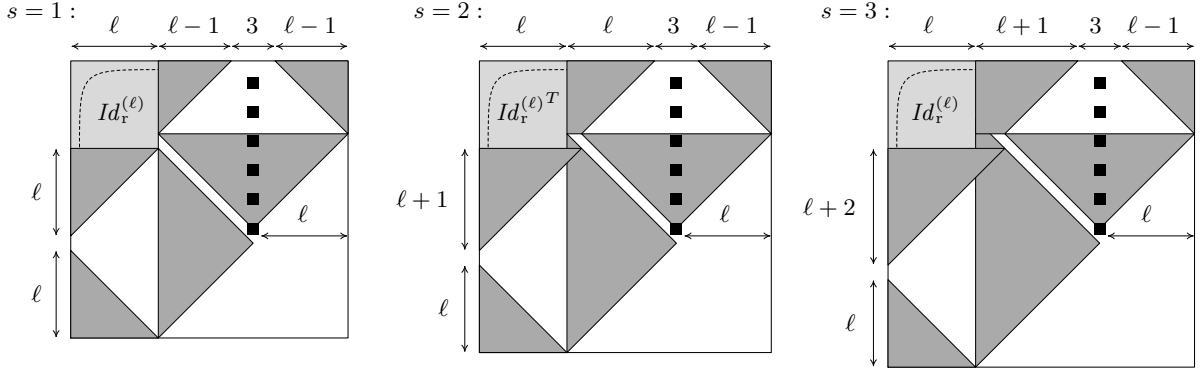}}
\put(0.1,2.5){\makebox[0pt][c]{$s=1:$}}
\put(0.6,2.4){\makebox[0pt][c]{$\ell$}}
\put(1.15,2.4){\makebox[0pt][c]{$\ell-1$}}
\put(1.55,2.4){\makebox[0pt][c]{$3$}}
\put(1.95,2.4){\makebox[0pt][c]{$\ell-1$}}
\put(1.9,1.1){\makebox[0pt][c]{$\ell$}}
\put(0.1,1.25){\makebox[0pt][r]{$\ell$}}
\put(0.1,0.55){\makebox[0pt][r]{$\ell$}}
\put(0.65,1.8){\makebox[0pt][c]{$\idr^{(\ell)}$}}
\put(2.9,2.5){\makebox[0pt][c]{$s=2:$}}
\put(3.4,2.4){\makebox[0pt][c]{$\ell$}}
\put(4.00,2.4){\makebox[0pt][c]{$\ell$}}
\put(4.45,2.4){\makebox[0pt][c]{$3$}}
\put(4.85,2.4){\makebox[0pt][c]{$\ell-1$}}
\put(4.8,1.1){\makebox[0pt][c]{$\ell$}}
\put(2.9,1.2){\makebox[0pt][r]{$\ell+1$}}
\put(2.9,0.45){\makebox[0pt][r]{$\ell$}}
\put(3.45,1.8){\makebox[0pt][c]{${\idr^{(\ell)}}^T$}}
\put(5.7,2.5){\makebox[0pt][c]{$s=3:$}}
\put(6.2,2.4){\makebox[0pt][c]{$\ell$}}
\put(6.85,2.4){\makebox[0pt][c]{$\ell+1$}}
\put(7.35,2.4){\makebox[0pt][c]{$3$}}
\put(7.75,2.4){\makebox[0pt][c]{$\ell-1$}}
\put(7.7,1.1){\makebox[0pt][c]{$\ell$}}
\put(5.7,1.15){\makebox[0pt][r]{$\ell+2$}}
\put(5.7,0.35){\makebox[0pt][r]{$\ell$}}
\put(6.25,1.8){\makebox[0pt][c]{$\idr^{(\ell)}$}}
\end{picture}
 \caption{\label{fig.idsManh}%
The non-recursive part of the identities $\idr^{(L)}$
(top-right corners) for $L=3 \ell + s$, and $s \in \{1,2,3\}$,
illustrating the set described by (\ref{eq.conds}).
For $s=2$, we show the transposed of $\idr$. 
Black and white stand respectively for $z_i=0$ and 1;
gray regiorns corresponds to chequered parts, starting with white on
the cells cutted by $\pi/4$-inclination lines.}
\end{figure*}

The statement of equations (\ref{eq.conds}) is graphically represented
in Figure~\ref{fig.idsManh}.

The understanding of the Creutz identity on
square portions of the pseudo-Manhattan lattice is completed by the
following theorem, relating the identity at side $2L+1$ to the one at
side $2L$. We encoded the identity at even sides in a $L \times L$
matrix $\idr^{(L)}$, exploiting the rotation symmetry, such that the
extended $2L \times 2L$ matrix has the property
\be
(\idr^{(L)})_{i,j} = (\idr^{(L)})_{2L+1-j,i}
\ef.
\ee
We can similarly encode the identity at odd sides in a structure of
almost $1/4$ of the volume, namely a $(L+1) \times (L+1)$ matrix 
$\idrodd^{(L)}$, using the fact that
\be
(\idrodd^{(L)})_{i,j} = (\idrodd^{(L)})_{2L+2-i,j} = (\idrodd^{(L)})_{i,2L+2-j}
\ef.
\ee

\begin{figure}
 \centering
\includegraphics[scale=0.48]{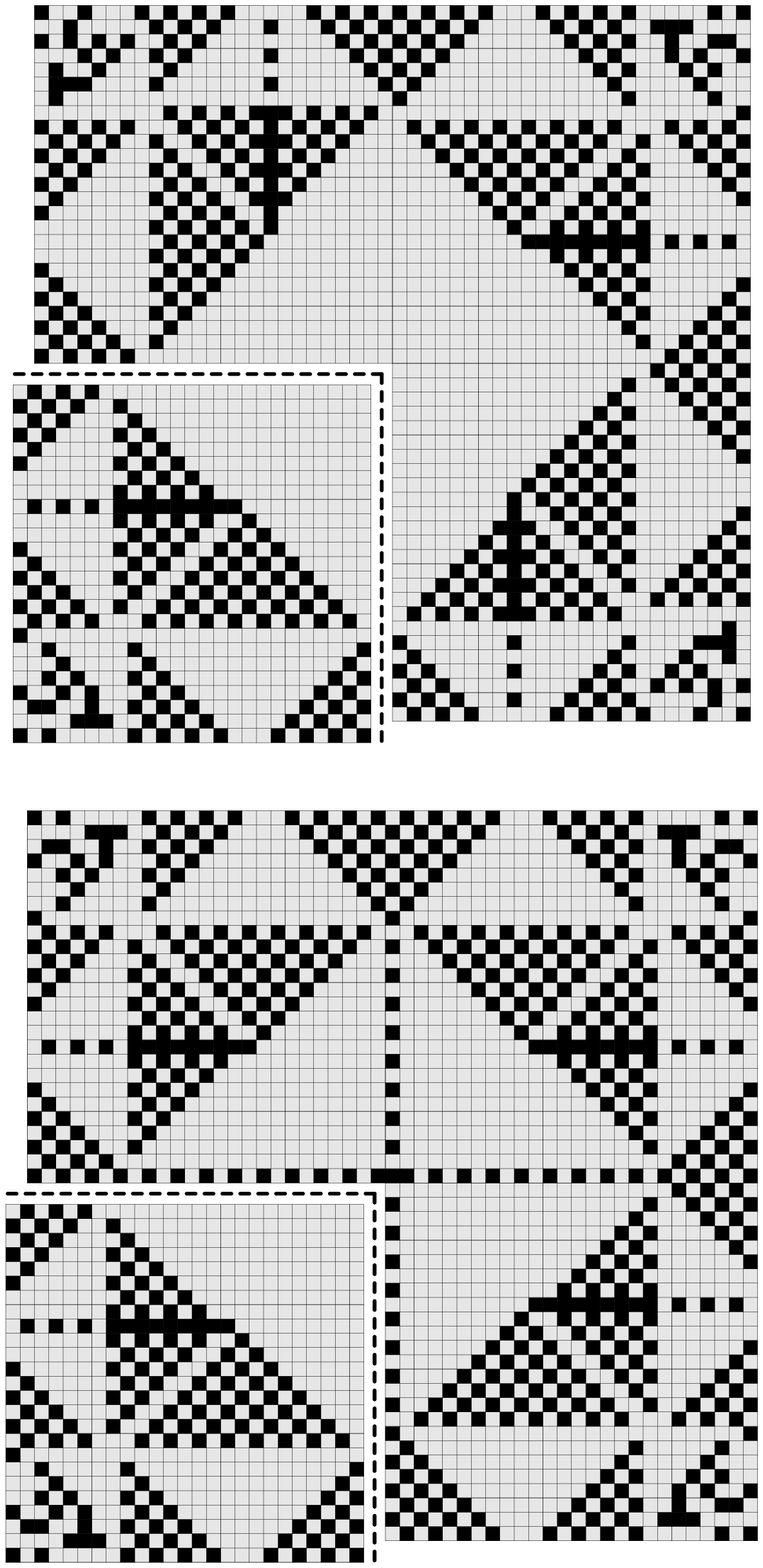}
 \caption{\label{figEvOd}%
The recurrent identity on the pseudo-Manhattan lattice of side $50$
(up) and $51$ (down), an example of how the identity on side $2L+1$ is
trivially deduced from the one on side $2L$.}
\end{figure}

\begin{theorem}
\label{theo.1b}
We have $(\idrodd^{(L)})_{i,j} = (\idr^{(L)})_{i,j}$ if $i,j\leq L$,
$(\idrodd^{(L)})_{L+1,j} = 0$ if $L-j$ is odd and 1 if it is even,
and, for $i \leq L$,
$(\idrodd^{(L)})_{i,L+1} = 0$ if $L-i$ is even and 1 if it is odd.
\end{theorem}
The statement of this theorem is illustrated in
Figure~\ref{figEvOd}.

The proof of Theorem \ref{theo.1b}, given
Theorem \ref{theorem1}, is easily achieved through arguments
completely analogous to the ones of \cite[sec.~7]{dhar_alg}.
We postpone the (harder) proof of Theorem \ref{theorem1} to the
discussion of the equivalence with Theorem \ref{theorem3} below.

There are some similarities and differences with the identities in the
BTW model. In that case, besides the evident height-2 square in the
middle, there arise some curvilinear triangolar structures, mostly
homogeneous at height 3, but some others ``texturized'', so that there
exists a variety of patterns which appear extensively at large sizes
(in the metaphor of CMYK offset printing, like the way in which
composite colours are produced!). A first attempt of classification of
these structures appears in \ref{srj}.  In our case, we only have
``black and white'', but, as four zeroes in a square are a forbidden
configuration (as well as many others too dense with zeroes), we can
not have extesive regions of zeroes (black, in our drawings of
fig.~\ref{fig.idsASMmanh}). A big square in the middle is still there,
rotated of $45$ degrees, while triangoloids are replaced by exact
``45-45-90'' right triangles in a ``texturized gray'' coming from a
chequered pattern. Indeed, to our knowledge, such a regular structure
as in theorem \ref{theorem1} was not deducible \emph{a priori} (in
particular, not before the publication of \cite{dharnew}), and our
initial motivation was to study the emergence of patterns in a
2-colour case.

The statement of Theorem \ref{theorem1} would suggest to look for
similar features also in the BTW model. It turns out that, while
in the directed case the $\lfloor (L-1)/3 \rfloor$ size is
\emph{fully} contained in the corner of the $L$ size, in the BTW the
$\lfloor L/2 \rfloor$ size is \emph{partially} contained in the corner
of the $L$ size, in an empirical way which strongly fluctuates with
$L$, but is in most cases more than $50 \%$ (cfr.~figure
\ref{fig.idsASM} for an example).

The heuristics in the case of the Manhattan lattice, under various
boundary prescriptions, are somewhat intermediate: the continuum-limit
configuration exists and coincides with the one for the
pseudo-Manhattan, but the $\lfloor (L-1)/3 \rfloor$ corners are only
partially reproduced by the size-$\ell$ identities.

How could one prove Theorem \ref{theorem1}? An elegant algebraic
property of the identity is that it is the only recurrent
configuration for which all the ``charges'' are zero (see
\cite{dhar_alg}, eqs.~(3.3) and (3.4)). More precisely, and in
a slightly different notation, given a whatever
ordering of the sites, and any site $i$, and calling $A_{i,j}$ the
minor $(i,j)$ of a matrix $A$, we have
\be
Q_j(C) := \sum_i 
z_i 
(-1)^{i+j} 
\det \Delta_{i,j} 
\ee
and $Q_j(\idr) = q_j \det \Delta$ with $q_j \in \mathbb{Z}$, with
$\idr$ being the only stable recurrent configuration with this
property.

A possible proof direction (that we do \emph{not} follow in this
paper) could have been as follows.
An algebraic restatement of the expression for the charges is achieved
in Grassmann calculus, through the introduction of a pair of
anticommuting variables $\psibar_i$, $\psi_i$ per site. Then, by
Grassmann Gaussian integration, we have that
\be
\label{eq.67857695}
(-1)^{i+j} \det \Delta_{i,j} 
= \int 
\mathcal{D}(\psi, \psibar) \,
\psibar_i \psi_j \,
e^{\psibar \Delta \psi}
\ee
where there is a contribution $b_k^- \psibar_k \psi_k$ in the
exponential for each site on the border, and a contribution
$(\psibar_h - \psibar_k) \psi_k$ if a particle falls into $h$ after a
toppling in $k$, i.e.~$\Delta_{hk}=-1$ (we are pedantic on this
because the asymmetry of $\Delta$ could create confusion on who's who
with $\psibar$ and $\psi$). 

So the configurations in the same equivalence class of the identity
are the only ones such that, for each $j$, the
``expectation value''
\be
\Big\langle \Big( \sum_i z_i \psibar_i \Big) \psi_j
\Big\rangle
:=
\frac{
\int \!
\mathcal{D}(\psi, \psibar) \,
(\sum_i z_i \psibar_i) \psi_j \,  
e^{\psibar \Delta \psi}
}
{
\int \!
\mathcal{D}(\psi, \psibar) \,
e^{\psibar \Delta \psi}
}
\ee
is integer-valued.

Furthermore, expressions as in the right hand side of (\ref{eq.67857695})
are related, through Kirchhoff theorem, to the
combinatorics of a collection of directed spanning trees, all rooted
on the boundary, with the exception of a single tree which instead
contains both $i$ and $j$, and the path on the tree from $i$ to $j$ is
directed consistently.

So, a possible approach by combinatorial bijections could be to prove
that, for any $j$, there is a suitable correspondence among the
forests as above, and a number $q_j$ of copies of the original
ensemble of rooted forests.

Such a task is easily performed, even for a generic (oriented)
graph, for what concerns the frame identity $\idf$, for which all
charges $q_j$ are 1. Unfortunately, for what concerns $\idr$, and with
an eye to the proof performed in the following section, it seems
difficult to pursue this project at least in the case of the
pseudo-Manhattan lattice on a square geometry. Indeed, if the
relaxation of $k_L \idf$ to $\idr$ requires $t_j$ topplings on site
$j$, it is easy to see that $q_j(\idr) = k_L - t_j$, and from the
explicit expressions for $k_L$ and the values of the topplings (the
latter are in the following Theorem \ref{theorem3}) we see that the
values of the charges $q_j$ are integers of order~$L^2$.

\section{Proof of the theorem}

Here we perform the direct proof of Theorem \ref{theorem1}. It is
fully constructive, a bit technical, and maybe not specially
illuminating for what concerns the algebraic aspects of the problem,
but still, it makes the job.

Now, in order to better exploit the geometry of our square lattice, we
label a site through a pair $ij$ denoting its coordinates.  We can
introduce the matrix $T_{ij}^{(L)}$, which tells how many topplings
site $ij$ performed in the relaxation of $k_L \idf$ into $\idr$.
Exploiting the rotational symmetry, we take it simply $L \times L$
instead of $2L \times 2L$,
with $(i,j)=(1,1)$ for the site at the corner, analogously to what we
have done for $(\idr^{(L)})_{ij}$.
Of course, although we call $T^{(L)}$ and $\idr^{(L)}$
``matrices'', they have a single site-index, and are indeed
vectors, for example, under the action of $\Delta$.

Clearly, $T$ is a restatement of $\idr$, as
\be
\label{eq.IT}
(\idr^{(L)})_{ij} = k_L\, b_{ij}^+
- \Delta_{ij, i'j'} \, T^{(L)}_{i'j'}
\ee
so that $\idr$ is determined from $T$, but also vice-versa, as
$\Delta$ is invertible. Actually, through the locality of $\Delta$,
one can avoid matrix inversion if one has some ``boundary condition''
information on $T$, and the exact expression for $\idr$, e.g.~if one
knows $T$ on two consecutive rows and two consecutive columns (and we
have a guess of this kind, as we discuss in the following).

The constraint that $(\idr)_{ij} \in \{0,1\}$ gives that $T$ is
locally a parabola with small curvature. Moreover,
in the regions corresponding to homogeneous portions
of $\idr$, $T$ must correspond to a discretized parabola through easy
formulas. The telescopic nature of 
$(\idr^{(L)}, \idr^{(\lfloor \frac{L-1}{3} \rfloor)}, \ldots)$
has its origin in an analogous statement for 
$(T^{(L)}, T^{(\lfloor \frac{L-1}{3} \rfloor)}, \ldots)$, and on the
fact that $k_L$ has a simple formula. As we will see, these facts are
easier to prove.

We start by defining a variation of $T$ which takes in account explicitly 
both the height-1 square in the middle of $\idr$,
and the spurious effects on the border.
Define $\tilde{k}(L) = \lfloor (L-1)(L+2)/4 \rfloor$ (which is
approximatively $k_L/2$), and introduce
\be
\hat{T}_{ij}^{(L)} := 
T_{ij}^{(L)} - \tilde{k}(L) b_{ij}^+
- \smbinom{L-i+1}{2} - \smbinom{L-j+1}{2}
\ef.
\ee
A first theorem is that
\begin{theorem}
\label{theorem2}
\begin{subequations}
\begin{align}
\label{eq.sulbordo}
\hat{T}_{iL}^{(L)} &= \hat{T}_{Lj}^{(L)} = 0
\ef;
\\
\hat{T}_{ij}^{(L)} &= 0
\qquad \qquad \qquad \textrm{for $i+j>L+\ell$;}
\\
\hat{T}_{ij}^{(L)} &\leq 0
\qquad \qquad \qquad \textrm{for all $i$, $j$.}
\end{align}
\end{subequations}
\end{theorem}
This implicitly restates the claim about the middle square of height 1
in $\idr$, and is in accord with the upper bound on the curvature of
$T$ given by $(\idr)_{ij} \leq 1$. With abuse of notations, we denote by
$\hat{T}_{ij}^{(\ell)}$ also the $L \times L$ matrix corresponding to
$\hat{T}_{ij}^{(\ell)}$ in the $\ell \times \ell$ corner with 
$i,j \leq \ell$,
and zero elsewhere. Then the rephrasing of the full 
Theorem \ref{theorem1} is
\begin{theorem}
\label{theorem3}
If $M_{ij}^{(L)} = 
-(\hat{T}_{L-i\;L-j}^{(L)} - \hat{T}_{L-i\;L-j}^{(\ell)})$
for $s=1,3$ and the transpose of the latter for $s=2$,
defining $\theta(n)=1$ for $n>0$ and 0 otherwise, and the 
``quadratic\,+\,parity'' function $q(n)$ on positive integers
\be
\begin{split}
q(n)
&=
\theta(n) \left(
n^2 + 2 n + \frac{1-(-1)^n}{2}
\right)
\\
&=
\left\{
\begin{array}{ll}
0 & n\leq 0\,; \\
(n+1)^2/4 & \textrm{$n$ positive odd;} \\
n(n+2)/4 & \textrm{$n$ positive even;}
\end{array}
\right.
\end{split}
\ee
then $M^{(L)}$ is a deterministic function, piecewise 
``qua\-drat\-ic\,+\,parity'' on a finite number of
triagular patches 
\be
\label{eq.M}
\begin{split}
M_{ij}^{(L)} &= 
\theta(i-L+\ell)
\Big[
- \smbinom{j}{2} 
- \max(0,j-\ell-1)
\\
& \!\!\!
+ q(j-i+L-\ell)
- q(j+i-L-\ell-4) 
\Big]
\\
&
+ \theta(j-L+\ell)
\Big[
- \smbinom{i}{2} 
-\theta(L-\ell-i+1)
\\
& \!\!\!
+ q(i-j+L-\ell-1)
+ q(i+j-L+\ell-2) 
\\
& \!\!\!
+ q(i+j-L-\ell-3) 
\Big]
\ef.
\end{split}
\ee
\end{theorem}
Clearly Theorem \ref{theorem2} is contained in Theorem \ref{theorem3},
just by direct inspection of the summands in (\ref{eq.M}). The
equivalence among Theorems \ref{theorem1} and \ref{theorem3} is
achieved still by direct inspection of (\ref{eq.M}), with the help of
some simple lemmas. First remark that the use of $-\hat{T}$ instead of
$T$ makes us work ``in false colours'', i.e.~effectively interchanges
$z$ into $1-z$ in $\idr$. Then, defining
$\nabla^2_x f(i,j) := f(i+1,j) + f(i-1,j) - 2 f(i,j)$, and analogously
$\nabla^2_y$ with $\pm 1$ on $j$,
we have that
\be
\nabla^2_{x,y} q(\pm i \pm j - a)
= 
\left\{
\begin{array}{ll}
1 & \pm i \pm j - a \geq 0 \textrm{\ and even;} \\
0 & \textrm{otherwise;}
\end{array}
\right.
\ee
and that $\nabla^2_x \max(0,j-a) = 1$ at $j=a$ only, while
$\nabla^2_y \max(0,j-a) = 0$ always (this reproduces the $\ell$ extra
zeroes out of the triangles depicted in figure~\ref{fig.idsManh}).

So, the explicit checks above can lead to the conclusion 
that equation (\ref{eq.IT}) holds at every $L$ for the expressions for
$T$ and $\idr$ given, in a special form, i.e.~subtracting the equation
for size $L$ with the one for size $\ell$, but on the $L \times L$
system. The latter is then easily related to the one on the 
$\ell \times \ell$ system, at the light of equation
(\ref{eq.sulbordo}), which allows to state that there are no different
contributions from having splitted the four quadrants of
$\hat{T}^{(\ell)}$. Thus, equation (\ref{eq.sulbordo}), guessing the
exact expression for $T_{ij}$ on the edges of each quadrant by mean of
a simple formula, is crucial to the possibility of having the
telescopic reconstruction procedure.

In other words, even without doing the tedious checks, we have seen
how, by a series of manipulations corresponding to the subtraction of
$\ell \times \ell$ corners to the $L \times L$ matrices, the proof is
restricted to the analysis of the ``deterministic'' part, with 
$i > \ell$ or $j > \ell$. As, in this case, all the involved functions
depend on $L$ through its congruence modulo $2$, or $3$, or $4$, it
suffices to check numerically the theorems for $12$ consecutive sizes
in order to have that the theorem must hold for all sizes. We did the
explicit check for sizes up to $L=64$.

The only thing that we need in order to conclude that the conjectured
expression $\idr$ corresponds to the identity is to prove that it is
indeed a recurrent configuration. Again, we do that in two steps, in
order to divide the behaviour on the self-similar $\ell \times \ell$
corner and on the deterministic part.

First, remark that in $\idr^{(L)}$ 
the internal border of the quadrant (the sites $ij$
with $i$ or $j=L$) is burnt in the Burning Test even without
exploiting the other mirror images. This is true because, at every $L$
and on both borders, we have a substructure of the form
\[
\includegraphics{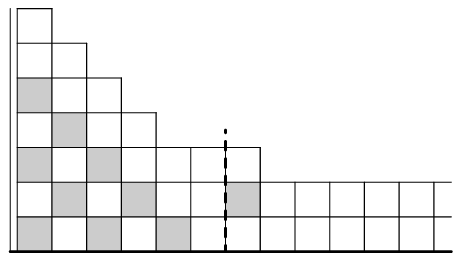}
\]
which is burnt through the following cascade (we put the burning
times, and denote with arrows from $x$ to $y$ a toppling occurring in
$x$ which triggles the toppling in $y$)
\[
\setlength{\unitlength}{12.07pt}
\begin{picture}(14,8)(-1.5,-1)
\put(-1.55,-0.85){\includegraphics[scale=1.2]{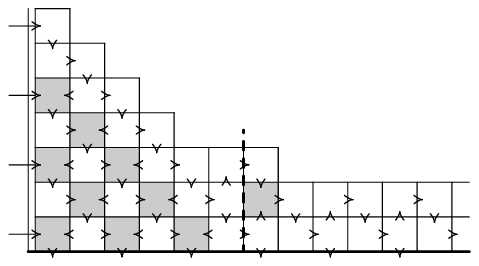}}
\put(0,0){\makebox[0pt][c]{$\scriptstyle{13}$}}
\put(2,0){\makebox[0pt][c]{$\scriptstyle{13}$}}
\put(4,0){\makebox[0pt][c]{$\scriptstyle{13}$}}
\put(6,0){\makebox[0pt][c]{$\scriptstyle{13}$}}

\put(1,0){\makebox[0pt][c]{$\scriptstyle{12}$}}
\put(3,0){\makebox[0pt][c]{$\scriptstyle{12}$}}
\put(5,0){\makebox[0pt][c]{$\scriptstyle{12}$}}

\put(0,1){\makebox[0pt][c]{$\scriptstyle{10}$}}
\put(2,1){\makebox[0pt][c]{$\scriptstyle{10}$}}
\put(4,1){\makebox[0pt][c]{$\scriptstyle{10}$}}
\put(6,1){\makebox[0pt][c]{$\scriptstyle{14}$}}
\put(1,1){\makebox[0pt][c]{$\scriptstyle{11}$}}
\put(3,1){\makebox[0pt][c]{$\scriptstyle{11}$}}
\put(5,1){\makebox[0pt][c]{$\scriptstyle{11}$}}

\put(7,0){\makebox[0pt][c]{$\scriptstyle{16}$}}
\put(8,0){\makebox[0pt][c]{$\scriptstyle{17}$}}
\put(9,0){\makebox[0pt][c]{$\scriptstyle{20}$}}
\put(10,0){\makebox[0pt][c]{$\scriptstyle{21}$}}
\put(7,1){\makebox[0pt][c]{$\scriptstyle{15}$}}
\put(8,1){\makebox[0pt][c]{$\scriptstyle{18}$}}
\put(9,1){\makebox[0pt][c]{$\scriptstyle{19}$}}
\put(10,1){\makebox[0pt][c]{$\scriptstyle{22}$}}

\put(0,2){\makebox[0pt][c]{$\scriptstyle{9}$}}
\put(2,2){\makebox[0pt][c]{$\scriptstyle{9}$}}
\put(4,2){\makebox[0pt][c]{$\scriptstyle{9}$}}
\put(6,2){\makebox[0pt][c]{$\scriptstyle{13}$}}
\put(1,2){\makebox[0pt][c]{$\scriptstyle{8}$}}
\put(3,2){\makebox[0pt][c]{$\scriptstyle{8}$}}
\put(5,2){\makebox[0pt][c]{$\scriptstyle{12}$}}

\put(0,3){\makebox[0pt][c]{$\scriptstyle{6}$}}
\put(2,3){\makebox[0pt][c]{$\scriptstyle{6}$}}
\put(1,3){\makebox[0pt][c]{$\scriptstyle{7}$}}
\put(3,3){\makebox[0pt][c]{$\scriptstyle{7}$}}

\put(0,4){\makebox[0pt][c]{$\scriptstyle{5}$}}
\put(2,4){\makebox[0pt][c]{$\scriptstyle{5}$}}
\put(1,4){\makebox[0pt][c]{$\scriptstyle{4}$}}

\put(0,5){\makebox[0pt][c]{$\scriptstyle{2}$}}
\put(1,5){\makebox[0pt][c]{$\scriptstyle{3}$}}

\put(0,6){\makebox[0pt][c]{$\scriptstyle{1}$}}

\put(11,0){\makebox[0pt][c]{$\scriptstyle{\cdots}$}}
\put(11,1){\makebox[0pt][c]{$\scriptstyle{\cdots}$}}
\end{picture}
\]
We choose to show just an example, instead of giving the explicit
formula, but it should be clear from the pictures above that a
general-$L$ regular procedure exists.

So, $\idr$ satisfies the burning test if and only if the deterministic
part of $\idr$ (the $L \times L$ square minus the $\ell \times \ell$
corner) satisfies the burning test with the border of the corner
having a $b_{ij}=1$ every two sites. This check must be performed only
on the deterministic part, and again is done in a straightforward way,
or, with conceptual economy, implied by the explicit numerical check
on 6 consecutive sizes. This completes the proof of all the three
theorems. \hfill $\square$

\section{\label{sec.manh}Manhattan Lattice}

\begin{figure}
 \centering
\includegraphics[scale=0.48]{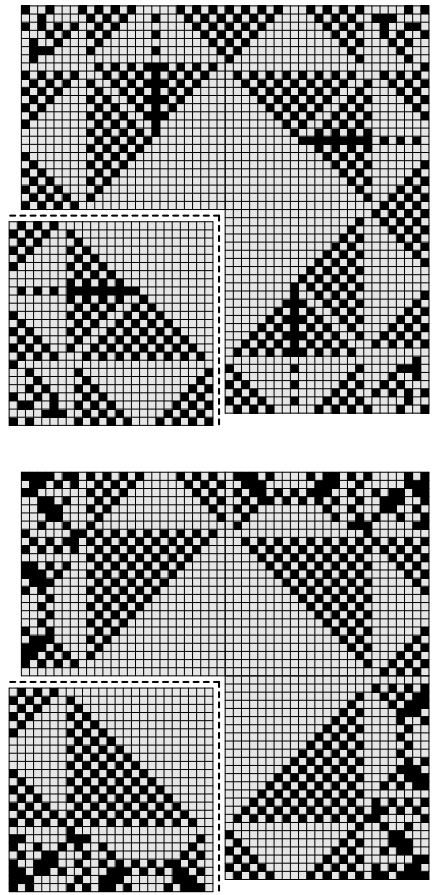}
 \caption{\label{figMpM}%
The recurrent identity on a portion of side $50$ of the square
lattice, with pseudo-Manhattan (up) and Manhattan (down) orientation.}
\end{figure}

For the Manhattan lattice, we performed an investigation similar to
the one above for the pseudo-Manhattan one, although, as we already
said, the motivations are less strong, as this lattice is not 
planar alternating.
A small-size example of Creutz identity, compared to the
pseudo-Manhattan one, is shown in Figure~\ref{figMpM}.

The numerics gave positive and negative results.
The positive results concern the filling number $k_L$, that,
according to extensive tests (up to $L \simeq 100$) we
conjecture to be
\be
k_L =
\left\{
\begin{array}{ll}
\frac{1}{4} \, L(L + 2)   & \textrm{$L$ even} \\
\frac{1}{4} \, (L + 1)^2  & \textrm{$L$ odd}
\end{array}
\right.
\ee
Furthermore, the whole quadrant except for the $\ell \times (\ell-1)$
corner seems to be deterministically described by a set of rules
analogous to equations (\ref{eq.conds}) 
(again, $\ell=\lfloor (L+1)/3 \rfloor$), and, as in (\ref{eq.conds}),
depending from the congruence of $L$ modulo 3, and a transposition
involved if $s=2$.

The negative result is that the telescopic exact self-similarity
between side $L$ and side $\ell$ seems to be lost in this case. As we
said, this property relies crucially on the simplicity of the toppling
matrix on the boundary of the quadrants, which seems to be an
accidental fact of the pseudo-Manhattan lattice, and has few chances
to show any universality. For this reason we did not attempt to state
and prove any theorem in the fashion of Theorem~\ref{theorem1} in this case.


\section{Conclusions}

We have studied, numerically and analytically, the shape of the Creutz
identity sandpile configurations, for variants of the ASM, with
directed edges on a square lattice (pseudo-Manhattan and Manhattan),
and square geometry.  An original motivation for this study was the
fact that heights are valued in $\{0,1\}$ (while the original BTW
sandpile has heights in $\{0,1,2,3\}$), and we conjectured that this
could led to simplifications. The results have been even simpler than
expected, and qualitatively different from the ones in the BTW model.

In the BTW model, the exact configuration seems to be unpredictable:
although some general ``coarse-grained'' triangoloid shapes seem to
have a definite large-volume limit, similar in the square geometry and
in the one rotated by $\pi/4$, here and there perturbations arise in
the configuration, along lines and of a width of order 1 in lattice
spacing. We discuss the role of these structures in various aspects of
the ASM in a forthcoming paper~\cite{us_pseudoprop}.

The triangoloids have precise shapes depending from their position in
the geometry, and are smaller and smaller towards the corners.
Understanding analytically at least the limit shape (i.e.~neglecting
all the sub-extensive perturbations of the regular-pattern regions) is
a task, at our knowledge, still not completed, although some first
important results have been obtained in
\cite{srj}, 
and further
achievements in this direction have come with the work of Levine and
Peres \cite{pereslevine, levinephd}, both in the similar context of
understanding the relaxation of a large pile in a single site (see in
particular the image at page 10 of Levine thesis, and the one at 

\noindent
\begin{tabular}{l}
{\tt http://math.berkeley.edu/}
\\
\quad {\tt \tildina levine/gallery/invertedsandpile1m10x.png},
\end{tabular}

\noindent
and, for the directed model, the one at
page 20 of

\noindent
\begin{tabular}{l}
{\tt newton.kias.re.kr/} \\
\quad {\tt \tildina nspcs08/Presentation/Dhar.pdf} ).
\end{tabular}



\begin{figure}
\setlength{\unitlength}{40pt}
\begin{picture}(5,9.2)
\put(0,0){\includegraphics[scale=0.79]{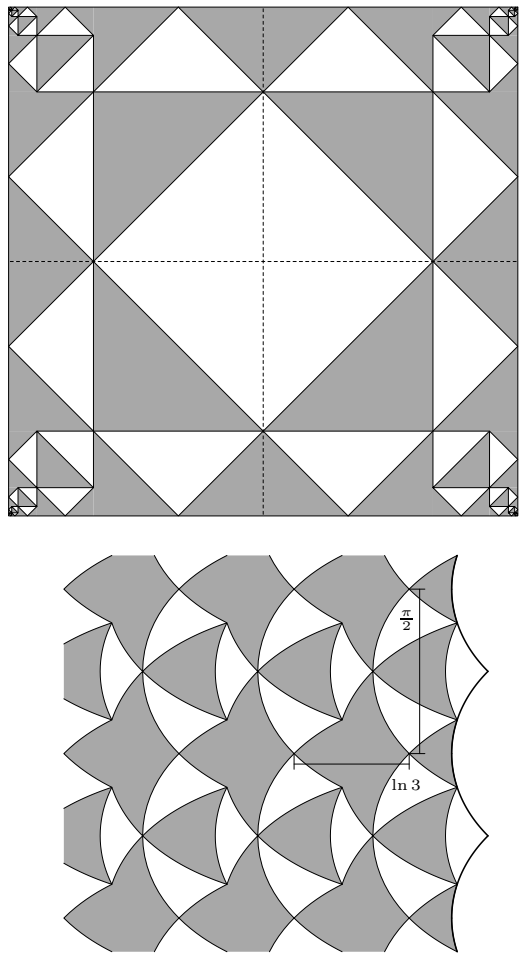}}
\put(3.72,1.67){$\scriptstyle{\ln 3}$}
\put(3.78,3.25){$\scriptstyle{\frac{\pi}{2}}$}
\end{picture}
 \caption{\label{fig.frac}%
Up: the limit Creutz Identity configuration on our Manhattan and
pseudo-Manhattan lattices. White regions correspond to have height $1$
almost everywhere. Gray regions correspond to have height $0$ and $1$
in a chequered pattern almost everywhere. Down: the image of a
quadrant under the map $z \to \ln z$.}
\end{figure}

In our Manhattan-like lattices on square geometry, however, we show
how the situation is much simpler, and drastically different.
Triangoloids are replaced by exact triangles, all of the same shape
(namely, shaped as half-squares), and with straight sides. All the
sides of the triangles are a fraction $\smfrac{1}{2} 3^{-k}$ of the
side of the lattice (in the limit), where the integer $k$ is a
``generation'' index depending on how near to a corner we are, and
indeed each quadrant of the configuration is self-similar under
scaling of a factor $1/3$. The corresponding ``infinite-volume'' limit
configuration is depicted in Figure~\ref{fig.frac}. A restatement of
the self-similarity structure, in a language resembling the $z \to
1/z^2$ conformal transformation in Ostojic \cite{srj} and Levine and
Peres \cite{pereslevine}, is the fact that, under the map $z \to \ln
z$, a quadrant of the identity (centered at the corner) is mapped in a
quasi--doubly-periodic structure.

Also the filling numbers (i.e.~the minimal number of frame identities
relaxing to the recurrent identity) have simple parabolic formulas,
while in the original BTW model a parabola is not exact, but only a
good fitting formula.

These features reach the extreme consequences in the pseudo-Manhattan
lattice, where the exact configuration at some size is
deterministically obtained, through a ratio-$1/3$ telescopic formula.
These facts are not only shown numerically, but also proven directly
in a combinatorial way.

\section*{Acknowledgements}

We are grateful to an anonymous referee for many useful comments, and
in particular for signaling us a very recent paper by Dhar and
collaborators \cite{dharnew} in which the same directed variants of
the ASM as in our paper are considered, for the related problem of
understanding the relaxation of a large pile of sand in a single
site. This is a further motivation for our work, and is quite likely
to be a precious reference for future research.


\end{document}